\documentclass[pre,aps,preprint,showpacs]{revtex4} 
\usepackage{revsymb,amsmath}
\usepackage{graphicx,psfrag,subfigure}

\newcommand{\goodgap}{%
  \hspace{\subfigcapskip}}
\begin{document}
\author{Steffen Trimper, Knud Zabrocki}
\affiliation{Fachbereich Physik, Martin-Luther-Universit\"at,D-06099 Halle Germany}
\email{trimper@physik.uni-halle.de}
\title{Memory Driven Pattern Formation}
\date{\today }

\begin{abstract}

The diffusion equation is extended by including spatial-temporal memory in such a manner that the   
conservation of the concentration is maintained. The additional memory term gives rise to the formation of 
non-trivial stationary solutions. The steady state pattern in an infinite domain is driven by a competition 
between conventional particle current and a feedback current. We give a general criteria for the existence 
of a non-trivial stationary state. The applicability of the model is tested in case of a strongly localized, time independent memory kernel. The resulting evolution equation is exactly solvable in arbitrary dimensions and 
the analytical solutions are compared with numerical simulations. When the memory term offers an spatially 
decaying behavior, we find also the  exact stationary solution in form of a screened potential.     

\end{abstract}
\maketitle

\section{Introduction}
There is an increasing effort in including feedback-couplings into the underlying evolution equations.  
Such memory-controlled effects should be a further unifying feature of complex physical \cite{6a,6b} 
as well as biological systems \cite{6c} far from equilibrium. Whereas most of the papers are addressed to 
a purely time dependent but homogeneous memory, the present one gives an extension to spatially dependent  
processes. Such systems, however without memory effects, are widely discussed in physics \cite{kas} and 
biology \cite{mur}. Based on our recent studies of several evolution models \cite{bst,st1,zab,zab1}, where 
such non-Markovian memory effects had been enclosed, and on the analysis of chemical reactions \cite{trizab}, 
we focus now on the diffusive behavior with memory couplings. These effects have an essential influence  
on the long time behavior, but they have also an impact on the dynamics in an intermediate time regime. As 
the consequence of the interplay between conventional diffusive transport and feedback we demonstrate the 
occurrence of non-trivial stationary solutions forming patterns.\\ 
\noindent Our analysis should be grouped into the continuing interest in feedback couplings. A memory dominated 
behavior is well established in analyzing the freezing 
processes in under-cooled liquids \cite{l,goe}, where the underlying mathematical representation is based 
on a projector formalism proposed by Mori \cite{mori}. As the result of the projection procedure the irrelevant 
degrees of freedom, fluctuating on short time and short length scales, contribute to the time evolution of the  
relevant degrees of freedom as well as by instantaneous and by delay-controlled terms. Since the projection formalism 
is not restricted to selected systems, the modification of the evolution equation due to memory effects is rather 
general. Here the relevant quantity may be the concentration of certain species or the probability density 
for a particle. The crucial influence of memory effects can be illustrated by considering a single particle moving 
in a disordered environment. Due to the strong disorder each member of an equally prepared ensemble makes  
experience by its own surrounding, which is even modified by the random walk of the single particle itself. 
Such a behavior can be described by a non-linear Fokker-Planck equation with an additional memory term as had been 
demonstrated in \cite{ss}. The analytical results, based on a one-loop renormalization group analysis, is in 
accordance with numerical simulations \cite{bst} and further analytical studies \cite{st1}. Whereas in that 
approach the memory effects are believed to be originated by the inherent non-linear interaction of the many particle system themselves, i. e. the time scale of the memory is determined by the relevant variable itself, there is a 
broad class of models which are subjected to external delay effects \cite{o,g,fe}; for a survey and applications in biology, see \cite{mathbio}. Even in traffic models such memory effects has been discussed \cite{he}.\\ 
Recently \cite{fk} memory effects in correlated anisotropic diffusion are studied in nanoporous crystalline solids. Likewise the effects of transport memory are discussed within the Fisher`s equation \cite{ah}, also applicable for bacterial population dynamics \cite{kk}. There appear a non-linear damping and traveling wave solutions \cite{abk}. The 
transport with memory, depending on the survivability of a population, is analyzed in \cite{h}.\\    
Regarding the large variety of systems with feedback couplings it seems to be worth to study simple models, which still include the dynamical features of evolution models as conservation of the relevant quantity $p(\vec r, t)$ and moreover, 
a time-delayed coupling. In the present paper we emphasize in an analytical solvable model with spatio-temporal 
feedback coupling. The generic behavior of the system may be changed by those additional delay effects. In particular, 
we discuss the transport behavior which is realized after a sufficient accumulation time and after cumulating particles 
within a spatial region. We show that the processes are affected by additional spatial and temporal 
correlations leading to non-trivial steady state solutions. 

\section{Model}

In case the transport is realized after a spatial-temporal accumulation process the time evolution 
of the probability could also depend on the history of the sample to which 
it belongs, i.e. the changing rate of the probability should be influenced by the changing rate in the past. 
Thus the evolution of the probability $p(\vec r, t)$ has to be supplemented by memory terms. Such a term models, 
for example the way on which a seed probability at a certain time $t'$ had been accumulated by a 
delayed transport mechanism, originated by the surrounded environment of the particle. In general, the changing 
rate of $p$ at time $t$ is also determined by the accumulation rate at a former time 
$t' < t$. In between, i.e. within the interval $\tau = t - t'$, the particles are enriched while 
changing the probability at $t'$. Regardless that process the available amount of particles at time $t$ is 
governed by an instantaneous transport term as well as by the changing rate at former times $t'$. 
Consequently the conventional diffusion equation should be modified according to \cite{spie}  
\begin{equation}
\partial _{t}p(\vec r,t) = \mathcal{M}(\vec r,t; p, \nabla p)  
+\int\limits_{0}^{t}dt'\int\limits_{-\infty }^{\infty }d^dr^{\prime}
\mathcal{K}(\vec r- \vec r^{\prime},t-t'; p, \nabla p)\mathcal{L}(\vec r^{\prime}, t'; p,\nabla p)\,.
\label{2}
\end{equation}
This equation is of convolution type. Whereas the operator $\mathcal{M}$ characterizes the instantaneous and 
local processes, both operators $\mathcal{K}$ and $\mathcal{L}$ are responsible for the delayed processes. In 
general the operators $\mathcal{M}, \mathcal{K}$ and $\mathcal{L}$ may be also non-linear in $p(\vec r, t)$ and 
$\nabla p(\vec r,t)$. They have to be specified according to the physical situation in mind. In particular 
we show that the operators are restricted when $p(\vec r,t)$ is assumed to be conserved. 
             
\subsection{Conservation}
In this section we specify the model, defined by Eq.~(\ref{2}), under the assumption that the basic 
quantity $p(\vec r, t)$ is conserved. This condition is fulfilled by 
\begin{equation}
\frac{d\,P(t)}{dt} = \frac{d}{dt} \int_{-\infty}^{+\infty} d^{d}r p(\vec r, t) = 0\quad .
\label{ev1}
\end{equation} 
To preserve $p$ the instantaneous term $\mathcal{M}$ has to be related to a current, e.g. 
$\mathcal{M}\propto \nabla \cdot \vec j $. Choosing natural boundary conditions, i.e. $p=0$ at the boundary, we 
find after Laplace transformation $P(t) \to P(z)$
\begin{equation}
z P(z) - P_0 = \hat{K}(z) \hat{L}(z) \quad\mbox{with}\quad \hat{K}(z) = \int d^dr\, \mathcal{K}(\vec r, z),\qquad 
\hat{L}(z) = \int d^dr\, \mathcal{L}(\vec r, z)\,,
\label{ev2}
\end{equation}  
with $P_0 = P(t=0)$. For an arbitrary polynomial kernel $\hat{K}$ the conservation law is in general not 
fulfilled provided the operator $\mathcal{L}$ is simply defined by $\mathcal{L} \equiv - \partial_t p(\vec r, t)$ 
(the minus sign is only for convention). Making this ansatz we conclude from Eq.~(\ref{ev2}) 
$$
[z P(z) - P_0][\hat{K}(z) + 1] = 0
$$
and consequently the conservation law is guaranteed. Physically, the assumption for $\mathcal{L}$ means, that we 
take into account a coupling of the rates, e.g. the evolution at the observation time $t$ is directly coupled to the changing rate at $t' < t$. Processes in the past will be permanently "reevaluated" at present time. In doing so the 
memory kernel gives rise to a coupling of the time scales. In the vicinity of the upper limit $t^{\prime} \simeq t$ the memory term reads $\mathcal{K}[\vec r, 0;\,p(\vec r, 0)]\partial_t p(\vec r, t)$, i.e. a instantaneous change at the observation time $t$ is coupled to the value of $p$ at the initial time $t = 0$. Therefore, the very past is related 
to the instantaneous value of $p(\vec r, t)$. In the opposite case, at the 
lower limit $t^{\prime} \simeq 0$, the change of the quantity $p(\vec r, t)$ near to the initial value 
$\partial_{t^{\prime}}\,p(\vec r, t^{\prime} = 0)$ is directly coupled to the instantaneous value $p(\vec r, t)$ via 
the kernel. In such a manner the memory part represents a weighted coupling of the behavior at the initial time and the observation time. Due to that coupling of the rates the long-time behavior of the system will be modified. 
One reason for that could be that the moving particle is embedded into an environment of all the other particles of the system. Due to the mutual interaction, the particle is released after a certain waiting time $t - t'$. Especially in sufficiently complex diffusion-reaction systems the feedback and memory effects should be relevant. In such system additional degrees of freedom like in flexible macromolecules in melts or in concentrated solutions \cite{dg}, nematic elastomer \cite{af} or in biology \cite{mathbio}. 
In that context one is interested in the description of effluent reprocessing plants in systems with closed 
water circulation. A special ecosystem of aerobic and anaerobic microorganisms is evolved in the 
clarifiers of such systems due to natural immigration or due to additional allowance. The living conditions 
of the microorganisms are mutually associated via the exchange of intermediate catabolic products. 
Each change of the concentration of one species will be stored in the food chain and effects the evolution of these  
species at a later time. Furthermore, the partial mixing in the clarifiers by convection and diffusion processes
enlarges the effects over the whole system, so that the memory integral introduced in Eq.~(\ref{2}) includes both, 
the time and the spatial coordinates. This special example may be extended also to other complex biological, chemical
or engineering problems with various hidden degrees of freedom, which are able to influence the evolution of a selected component significantly, for instance by biological interaction with other species via the food chain or via biological competition. Such effects which are partially observable, could contribute to the memory term. 

\subsection{General properties}

\noindent Starting from Eq.~(\ref{2}) we can make some general statements for an arbitrary kernel 
$\mathcal{K}(\vec r, t; p, \nabla p)$ and the specialized term $\mathcal{L}(\vec r, t; p, \nabla p) = - \partial_t p(\vec r, t)$. As stressed before the last condition guarantees the conservation of $p(\vec r, t)$. After Fourier transformation 
with respect to the spatial coordinate and Laplace transformation with respect to the time, we get from Eq.~(\ref{2})
\begin{equation}
p(\vec k, z) = \frac{p_0(\vec k)}{z + \hat{D}(\vec{k}, z)\,k^2}\quad\text{with}\quad 
\hat{D}(\vec{k}, z) = \frac{D}{1 + \mathcal{K}({\vec{k}}, z)}\,,
\label{ev10}
\end{equation} 
where $\mathcal{K}(\vec k, z)$ is the Fourier-Laplace transformed kernel. Eq.~(\ref{ev10}) describes diffusion with a modified diffusion parameter $D \to \hat{D}(\vec r, t)$. Assuming a regular behavior of the 
kernel let us make the ansatz
\begin{equation}
\mathcal{K}(\vec k, z) = \frac{b(\vec k)}{z} + \Lambda (\vec k, z)\quad\text{with}
\quad\lim_{z \to 0}\Lambda ({\vec k}, z)= \text{finite}\quad .
\label{ev11}
\end{equation}
Provided the kernel reveals a finite stationary value $\lim\limits_{t \to \infty} \mathcal{K}(\vec k, t) \equiv b(\vec k)$, then $p(\vec r, t)$ yields a stationary solution, too. Inserting Eq.~(\ref{ev11}) in Eq.~(\ref{ev10}) we obtain
\begin{align}
p(\vec k, z) &= \frac{p_s(\vec k)}{z} + \varphi (\vec k, z)\nonumber\\
\text{with} \quad p_s(\vec k) &= \frac{p_0(\vec k) b(\vec k)}{b(\vec k) + D k^2}\quad\text{and}\quad
\varphi (\vec k, z) = \frac{p_0(\vec k)\,Dk^2\,[\,1 + \Lambda (\vec k, z)]}{[\,b(\vec k) + D k^2\,]
[\,z(1 + \Lambda (\vec k, z)) + b(\vec k) + Dk^2]}\,. 
\label{ev12}
\end{align} 
Summarizing the results we conclude that the model, following Eq.~(\ref{2}) for the conserved quantity $p(\vec r, t)$,    
gives rise to a non-trivial stationary solution $g(\vec k$), or equivalent $g(\vec r)$, if the Laplace transformed 
kernel $\mathcal{K}(\vec k, z)$ satisfies the condition
\begin{equation}
\lim_{z \to 0} z \mathcal{K}(\vec k, z) \neq 0\,.
\label{ev14a}
\end{equation}
This result is a generalization of a previous one derived for a homogeneous system \cite{trizab}.

\section{Diffusion with time independent memory kernel }

In the following let us regard the evolution equation 
\begin{equation}
\partial_t p(\vec r, t) = D\nabla^2p(\vec r, t) - \int_{-\infty}^{\infty} d^dx' \int_0^t dt'\, K(\vec r - \vec r', t - t')\,
\partial_{t'}p(\vec r',t')\,. 
\label{ev3}
\end{equation}
\noindent As discussed in the previous section the conservation of the quantity $p(\vec r,t)$ is guaranteed for an arbitrary kernel $K$. Different to a conventional evolution equation, in the present one $p(\vec r, t)$ is related to 
$p(\vec r, t')$ with $0< t'< t$. Moreover, Eq.~(\ref{ev3}) offers also a coupling between $\partial_t p(\vec r, t)$ and 
$\partial_{t'}p(\vec r, t')$. The mixing of time scales leads, above all, to a substantial modification in the long 
time limit. 

\subsection{Spatially localized memory}
\begin{figure}[!ht]\centering
\psfrag{c(x,t)}[][][0.6]{$p$}
\psfrag{ps(x)}[][][0.6]{$p_s$}
	\subfigure[Stationary solution]{\includegraphics[scale=0.25,angle=-90]{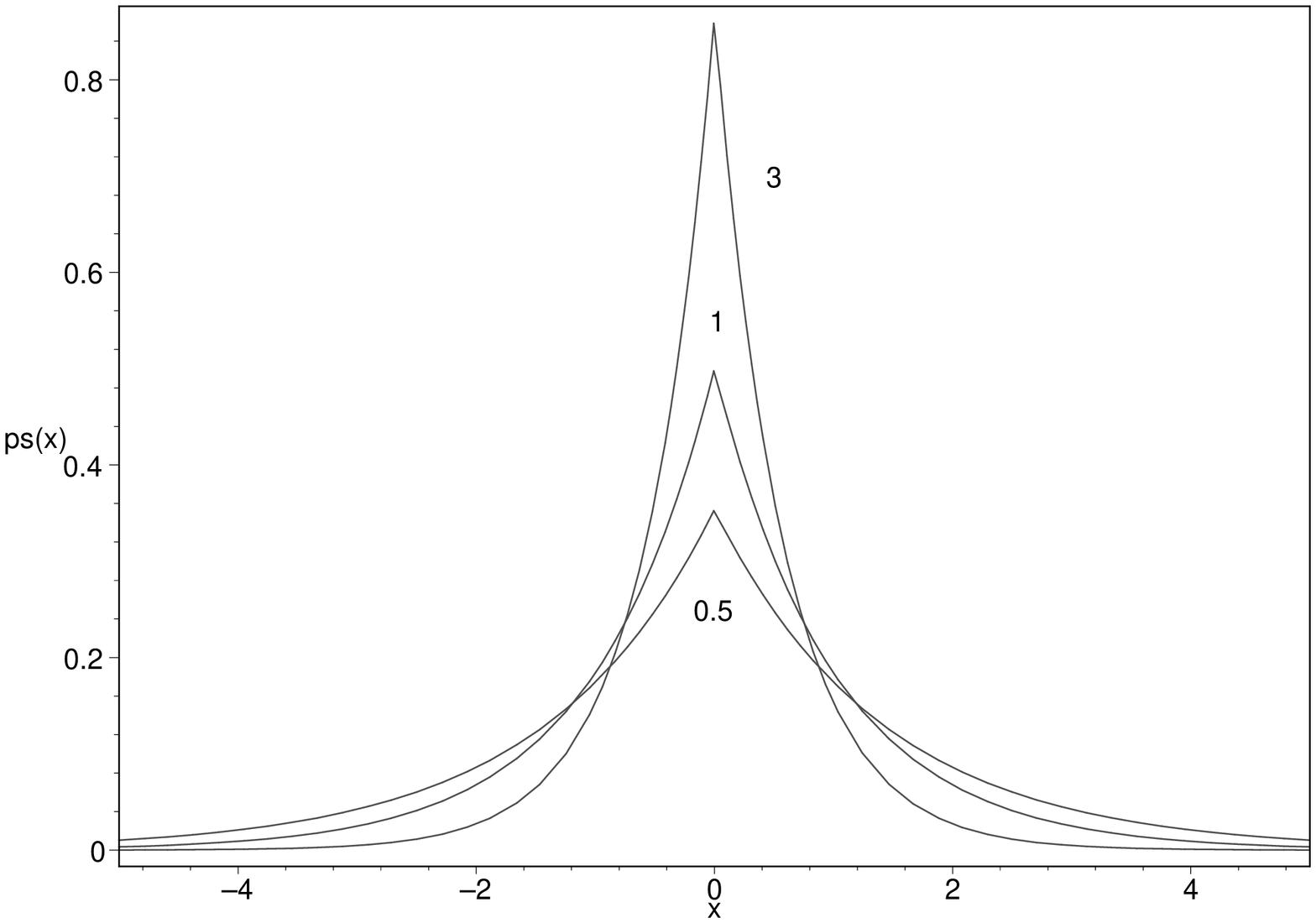}}
	\goodgap
	\subfigure[$\mu = 0.5$]{\includegraphics[scale=0.3,angle=-90]{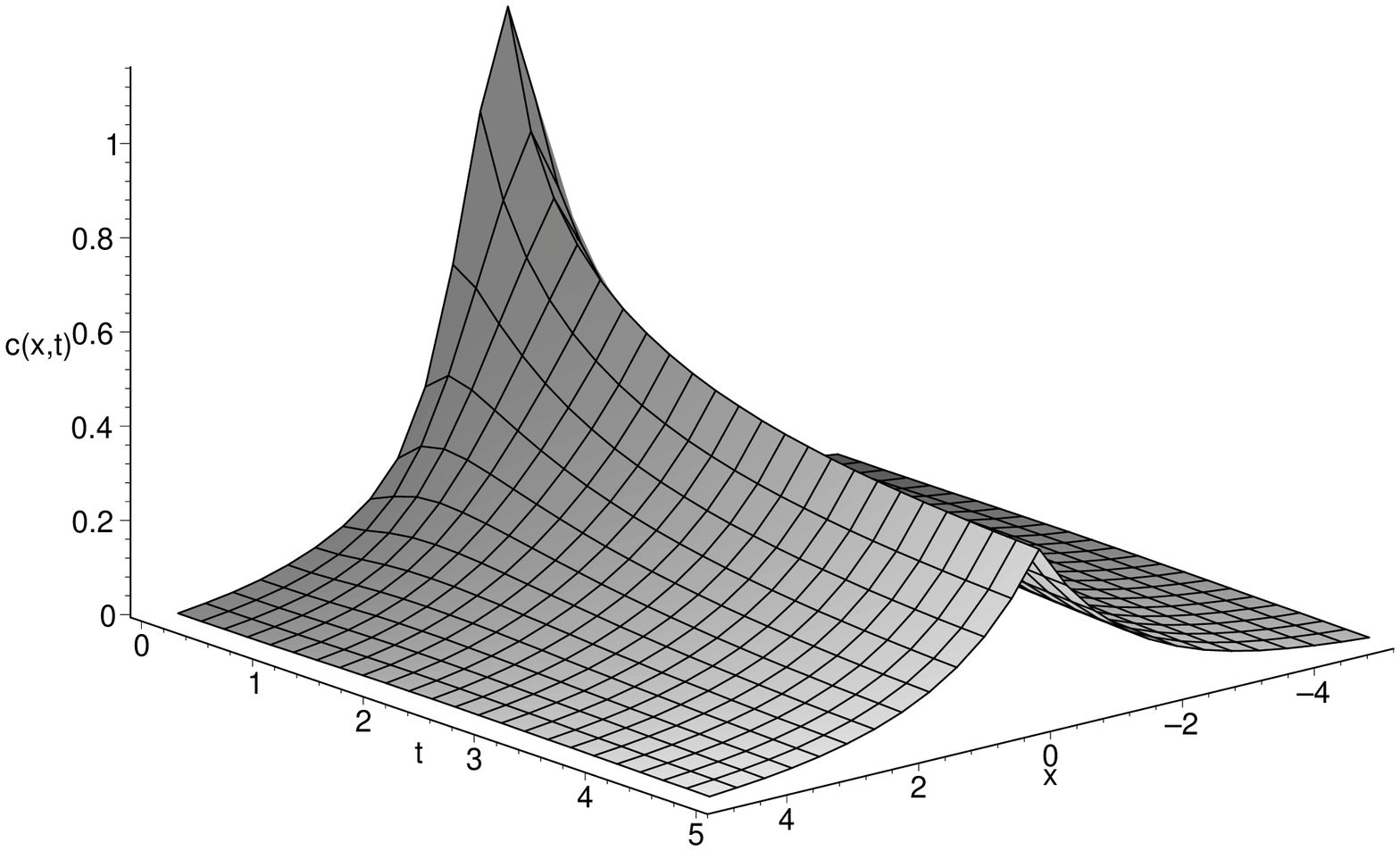}}\\
	\subfigure[$\mu = 1$]{\includegraphics[scale=0.3,angle=-90]{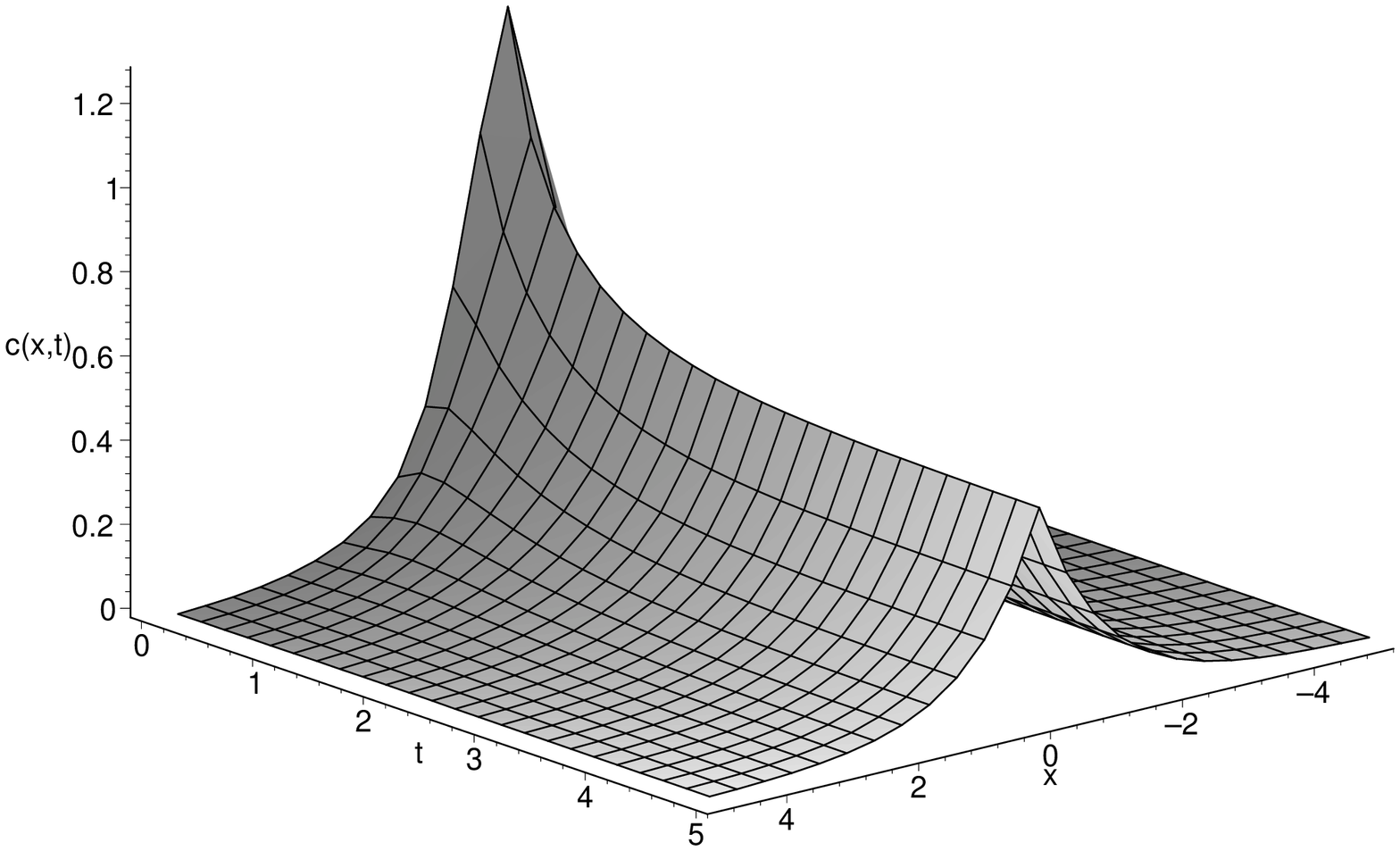}}
	\goodgap
	\subfigure[$\mu = 3$]{\includegraphics[scale=0.3,angle=-90]{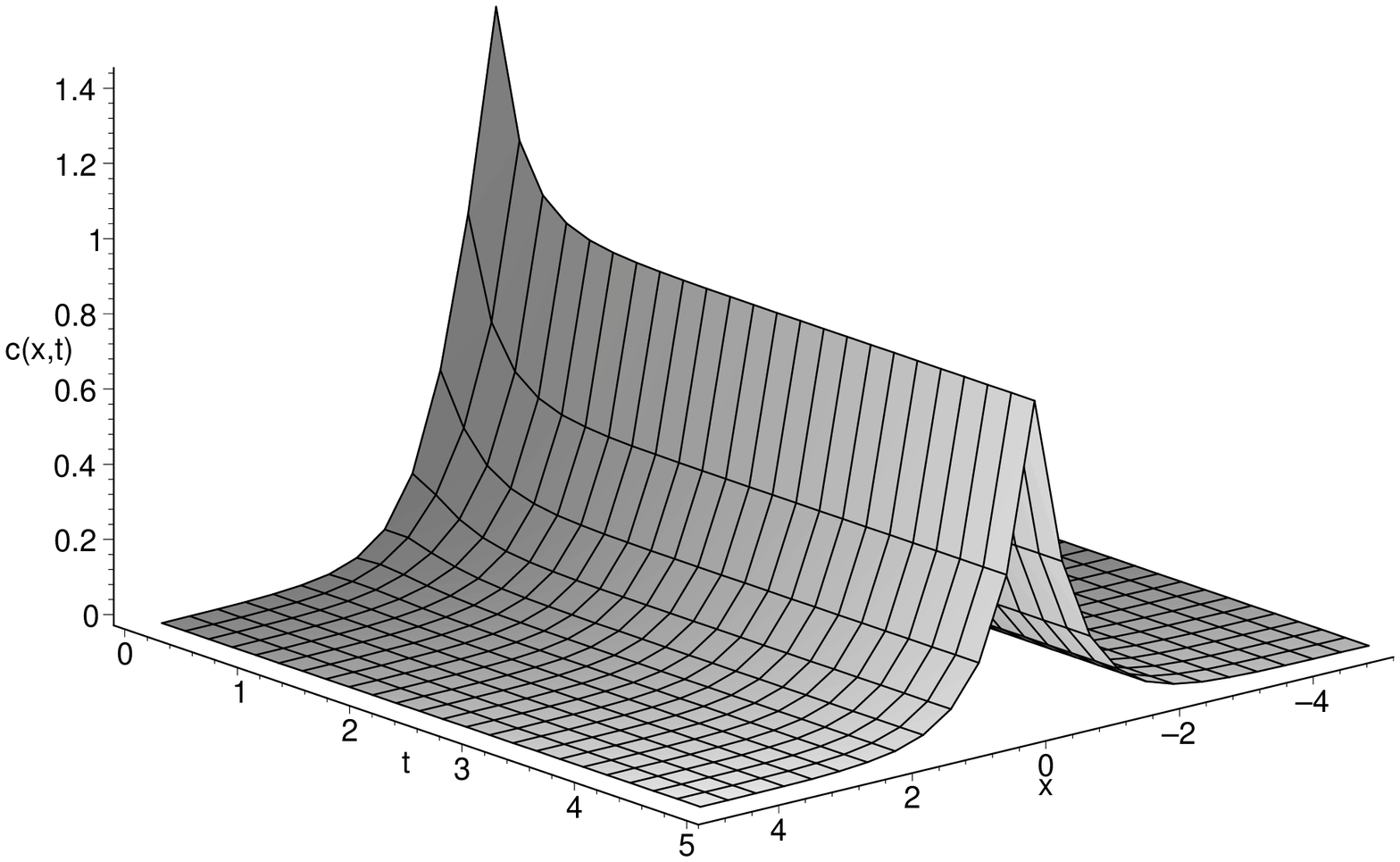}}\\
\caption{Stationary solution $p_s(x)$ and $p(x,t)$ in $d=1$ with $D=1$ and $p_0 = 1$ for different strength of memory 
$\mu $.} 
\label{Fig.1}
\end{figure}

As the simplest example we analyze a strictly spatially local, but time independent kernel
\begin{equation}
K(\vec r, t) = \mu\, \delta (\vec r)\, ,
\label{ev3a}
\end{equation}
where $\mu > 0$ is the strength of the memory. Note that in our approach spatial and temporal variables are 
decoupled. In case of a coupling between space and time variable one may observe different effects discussed elsewhere \cite{zts}. Inserting this kernel in Eq.~(\ref{ev3} we get
\begin{equation}
\partial_t p(\vec r, t) = D\nabla^2p(\vec r, t) - \mu [p(\vec r, t) - p_0(\vec r)] \quad\mbox{with}\quad p_0(\vec r) \equiv  p(\vec r, t =0)\quad .
\label{ev4}
\end{equation}
Such a memory term leads to a very strong coupling of the instantaneous value to the initial value 
because all previous time steps are weighted equally. There appears 
a memory induced feedback to the initial distribution as a driving force. Note that without the coupling to the 
initial value $p_0(\vec r)$, the particle would perform a diffusive motion and the probability density $p(\vec r,t)$ 
would decay on the time scale $\mu ^{-1}$. Mathematically the inhomogeneous linear equation may be solved analytically 
for arbitrary initial conditions and in any dimension: 
\begin{equation}
p(\vec r,t)=\text{e}^{-\mu\, t} \int_{-\infty}^{\infty} G (\vec r - \vec r\,', t) p_0 (\vec r\,')\, d^d\, r' 
 +\mu\, \text{e}^{-\mu\, t} \int_{-\infty}^{\infty}\,\int_{0}^{t} G(\vec r-\vec r\,' ,t-t')\, \text{e}^{\mu\, t'}\, p_0 (\vec r\,' ) dt' \,  d^d r' \,.
\label{gs}
\end{equation}
From this solution one can verify that a non-negative initial distribution $p_0(\vec r\,) \geq 0$ leads always 
to $p(\vec r, t) \geq 0$ provided $\mu > 0$. Eq.~\eqref{gs} enables us to calculate the second moment 
$s(t)$. It results
\begin{equation}
s(t)\equiv\int\limits_{-\infty}^{\infty} \vec{r}^{\,2} \, p(\vec r,t) \, d^d r =\frac{2\,d\, D\, \left(1-\text{e}^{-\mu\, t}\right)}{\mu}\, \int_{-\infty}^{\infty} p_0 (\vec r\,)\, d^d r + \int_{-\infty}^{\infty}\vec{r}^{\,2} \,  p_0 (\vec r\,)\, d^d r \,.
\end{equation}
Notice that for $\mu \to 0$ the last equation shows the conventional diffusive behavior.\\
Let us illustrate the behavior in the one dimensional case with the initial condition $p_0(x) = p_0 \delta (x)$. 
The exact solution reads
\begin{align}
p(x,t) &= \frac{p_0}{\sqrt{4\pi Dt}}e^{-(\mu t +\frac{x^2}{4Dt})} + 
\frac{p_0 \kappa }{4}\left\{e^{\kappa x}\left[\rm erf\left(\frac{x}{\sqrt{4Dt}}+\sqrt{\mu t}\right)-\rm sgn(x)\right]\right\} \nonumber\\
&\quad + \frac{p_0 \kappa }{4}\left\{e^{-\kappa x}\left[\rm erf\left(-\frac{x}{\sqrt{4Dt}}+\sqrt{\mu t}\right) - \rm sgn(-x)\right]\right\}\,,
\label{ev5}
\end{align}   
where $\rm erf (x)$ is the error function and $\kappa = \sqrt{\mu /D}$. The first part of Eq.~(\ref{ev5}) is the solution 
of the homogeneous equation showing a temporal decay with the time constant $\mu ^{-1}$. For different 
strength of the feedback coupling $\mu = \,0.5,\,1,\,3$ the solution $p(x,t)$ is presented in Fig.~(\ref{Fig.1}).  
The stronger the feedback strength $\mu$ the more pronounced are the deviations from the pure diffusive behavior. 
In particular, the system develops a non-trivial stationary solution whenever $\mu > 0$. Such a stationary solution 
is due the permanent coupling to the initial distribution. We find 
\begin{equation}
\lim_{t \to \infty} p(x,t) \equiv p_s(x) = \frac{p_0 \kappa }{2} e^{-\kappa \mid x \mid}\,.
\label{ev6}
\end{equation}
where $p_s(x)$ is also included into Fig.~(\ref{Fig.1}). Notice that in the limiting case 
$\kappa \to \infty$ the stationary solution tends to the initial distribution. Obviously our feedback offers a 
kind of shape memory. The stronger the memory is the stronger is the recognition of the initial state.\\ 
\noindent The result can be generalized to arbitrary dimensions $d$. For the stationary solution $p_s(\vec r)$ we obtain 
with $r = \mid \vec r \mid$
\begin{equation}
p_s(r) = \frac{p_0}{(2\pi )^{d/2}} \kappa ^{\frac{d+2}{2}} r^{-\frac{d-2}{2}}K_{\frac{d-2}{2}}(\kappa r)\,.
\label{ev7}
\end{equation} 
The modified Bessel function $K_{\nu}(x)$ can be expressed by standard functions for $d = 1, 3$ \cite{as}. In the two 
dimensional case the modified Bessel function offers a logarithmic behavior. 
The evolution equation for $p(\vec r, t)$ is integrable with
\begin{equation}
p({\vec r} ,t) = \frac{p_0}{(4\pi D t)^{d/2}} e^{-\left(\mu t + \frac{\vec r^2}{4Dt}\right)} + 
\mu p_0 e^{-\mu t}\int _0^t dt' e^{\mu t'}\frac{1}{[4\pi D(t - t')]^{d/2}} e^{-\frac{\vec r^2}{4D(t-t')}}\,.
\label{ev8}
\end{equation}
The last integral can be calculated exactly for $d=1$ with the result given in Eq.~(\ref{ev5}). 
Asymptotically we obtain
\begin{equation}
p({\vec r} ,t) \approx \frac{p_0}{(4\pi D t)^{d/2}} e^{-\left(\mu t + \frac{\vec r^2}{4Dt}\right)} 
+ \frac{p_0}{(2\pi )^{d/2}} \kappa ^{\frac{d+2}{2}} r^{-\frac{d-2}{2}}K_{\frac{d-2}{2}}(\kappa \mid \vec r \mid)\,.
\label{ev9}
\end{equation}
The analytical results are confirmed by numerical simulations.

\subsection{Gaussian and exponential initial conditions}

\begin{figure}[t]
\centering
\psfrag{c (x,t)}[][][0.6]{$p$}
\psfrag{ps(x)}[][][0.6]{$p_s$}
	\subfigure[Stationary solution]{\includegraphics[scale=0.25,angle=-90]{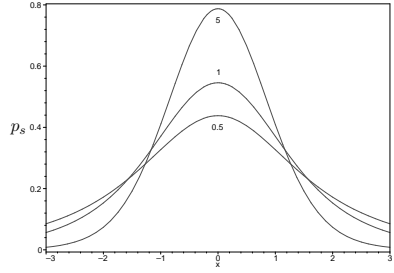}}
	\goodgap
	\subfigure[$\mu = 0.5$]{\includegraphics[scale=0.3,angle=-90]{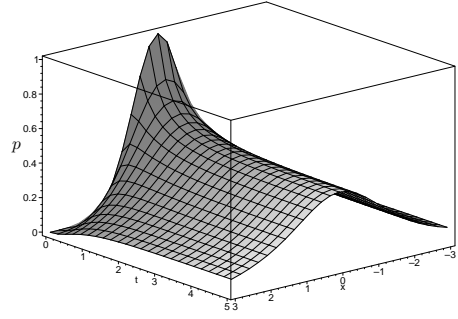}}\\
	\subfigure[$\mu = 1$]{\includegraphics[scale=0.3,angle=-90]{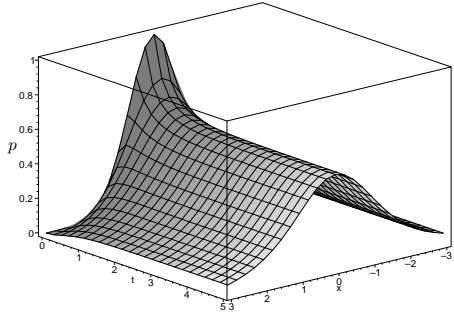}}
	\goodgap
	\subfigure[$\mu = 5$]{\includegraphics[scale=0.3,angle=-90]{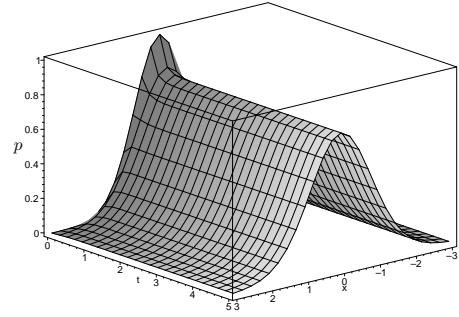}}\\
\caption{Stationary solution $p_s(x)$ and $p(x,t)$ for the Gaussian initial distribution with $D=1,\,d=1,\,p_0=1,\,\lambda = 1$ and 
different $\mu $.}
\label{Fig.2}
\end{figure}

\begin{figure}[t]
\centering
\psfrag{c (x,t)}[][][0.6]{$p$}
\psfrag{ps(x)}[][][0.6]{$p_s$}
	\subfigure[Stationary solution]{\includegraphics[scale=0.25,angle=-90]{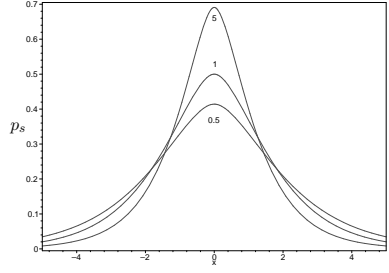}}
	\goodgap
	\subfigure[$\mu = 0.5$]{\includegraphics[scale=0.3,angle=-90]{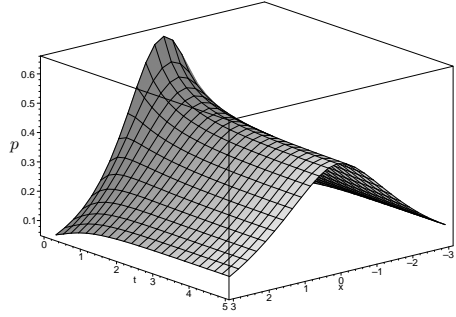}}\\
	\subfigure[$\mu = 1$]{\includegraphics[scale=0.3,angle=-90]{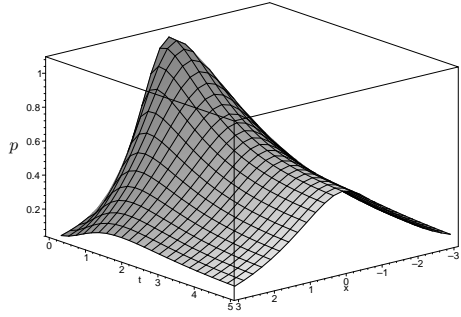}}
	\goodgap
	\subfigure[$\mu = 5$]{\includegraphics[scale=0.3,angle=-90]{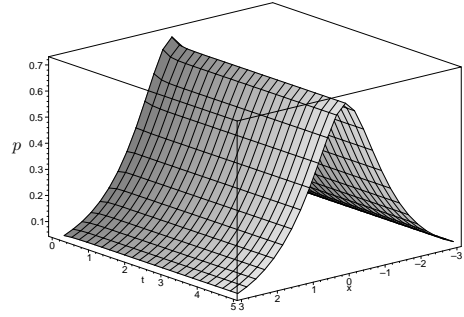}}\\
\caption{Stationary solution $p_s(x)$ and $p(x,t)$ for the exponential initial distribution with $D=1,\,d=1,\,p_0=1,\,\lambda = 1$ and 
different $\mu $.}
\label{Fig.3}
\end{figure}

For an arbitrary initial condition $p_0(\vec r)$ the stationary solution can be directly calculated by solving 
equation $\nabla ^2p_s(\vec r) = \kappa ^2 [\,p_s(\vec r)-p_0(\vec r)\,]$ with $\kappa ^2 = \mu /D$. It results in 
\begin{equation}
p_s(\vec r) = \frac{\kappa^{(d+2)/2}}{(2\pi)^{d/2}}
\int d^dx'\frac{p_0(\vec r')}{\mid \vec r - \vec r'\mid^{\frac{d-2}{2}}}\,
K_{\frac{d-2}{2}}(\kappa \mid \vec r - \vec r'\mid)\quad .
\label{ev9a}
\end{equation} 
In this subsection we study different realizations of the initial conditions. In case of a Gaussian initial distribution 
$p_0(\vec r) = p_0 e^{-\lambda x^2}$ the results for different strength of memory $\mu $ for both $p(x,t)$ as well as 
$p_s(x)$ are shown in Fig.~(\ref{Fig.2}). When the initial condition is an exponentially decaying function 
$p_0(\vec r) = p_0 e^{-\lambda \mid \vec r \mid}$, we find solutions presented in Fig.~(\ref{Fig.3}).  
\noindent Summarizing this section we have shown, that a simple feedback coupling lead to stable patterns 
which do not appear under a conventional diffusion term.

\section{Fading diffusion}

Now we regard as a further realization of Eq.~(\ref{2}) with a conserved $p(\vec r, t)$ a spatially decaying memory 
term by choosing the kernel in the form 
\begin{equation}
\mathcal{K}(\vec r, t) = \mu \exp(-\lambda \mid \vec r \mid)\,.
\label{exp}
\end{equation}
Performing Laplace transformation we get
$$
\mathcal{K}(\vec k, z) = \frac{A \mu \lambda }{z \,(\lambda^2 + k^2)^{(d+1)/2}}\quad\mbox{with}\quad 
A(d)=2^d(\pi )^{(d-1)/2}\Gamma \left(\frac{d+1}{2}\right)
$$
Inserting that result into Eq.~(\ref{ev12}) it results a non-trivial stationary solution, compare also Eq.~(\ref{ev12}), 
\begin{equation}
p_s(\vec k)= \frac{p_0(\vec k) A \mu \lambda }{A \mu \lambda + D k^2 [\lambda ^2 + k^2]^{\frac{d+1}{2}}}\,.
\label{exst}
\end{equation}
\noindent In the hydrodynamic limit $\vec k \to 0$ one finds the Fourier transformed stationary solution  
\begin{equation}
p_s(\vec r) = \frac{\kappa ^{d/2+1}}{(2\pi )^{d/2}}\int d^dx'\frac{p_0(\vec r')}{\mid \vec r -\vec r'\mid^{\frac{d-2}{2}}} 
K_{\frac{d-2}{2}}(\kappa \mid \vec r - \vec r'\mid)\quad\mbox{with}\quad \kappa =\sqrt{\frac{A\mu}{D \lambda ^d}}\,.
\label{exst1}
\end{equation} 
Using the asymptotic expansion for the modified Bessel functions \cite{as}, one concludes that the stationary solution offers a screened Coulomb potential decaying on a length scale $\kappa ^{-1}$. Note that the screening length is determined 
by the memory strength $\mu $ as well as by the characteristic length $\lambda ^{-1}$ of the spatial memory. The 
screening length increases when the memory strength $\mu $ decreases. Let us stress that the non-trivial steady state 
is originated by the competition between diffusion and memory and it is not driven by spatial boundary conditions. In 
that case there appears an additional length scale, the size of the sample $L$, which gives rise to further effects not 
discussed now.

\section{Conclusions}

In this paper we have extended the conventional diffusive transport by including non-Markovian 
memory terms within the frame of evolution equations. The additional terms are chosen in such a manner that the 
relevant variable $p(\vec r, t)$ is normalized. By this requirement the form of the memory term is restricted 
to a class where the changing rate of $p$ at the observation time $t$ is coupled to the changing rate of $p$ 
at a former time $t'$. Insofar the delay effect offers a long time memory due to $0 \leq t' \leq t$. Further 
the model exhibits also a long range memory because the kernel $\mathcal{K}$ depends on $\vec r - \vec r^{~\prime}$. 
Here we have considered several realizations of the memory kernel with different behavior. As a new result there 
occur non-trivial stationary solutions forming patterns. Such patterns are absent for conventional diffusion in an 
infinite domain without boundary conditions. The reason for such a new behavior is 
by means of the above mentioned coupling of the rates. This time accumulation 
is further accompanied by an additional spatial accumulation, the effect of which is comparable to the effect 
a long-range interaction forces and consequently the results are basically independent of the spatial  
dimensions in according to scaling arguments. These many-body effects are shown to change the asymptotic 
behavior drastically. Due to the feedback-coupling of a particle to its environment, a subsequent particle, 
undergoing a diffusive motion, gains information from a modified environment. As the result there appears 
patterns controlled by the memory strength.

\begin{acknowledgments} 
This work is supported by the DFG (SFB 418). We are indebted to Peter H\"anggi and Gunter Sch\"utz for 
valuable discussions. 
\end{acknowledgments}

\newpage

\end{document}